\documentclass[12pt,english]{article}

\usepackage{natbib}
\usepackage{url}
\makeatletter
\def\url@leostyle{%
    \def\UrlFont{\sf}}{\def\UrlFont{\small\ttfamily}}
\makeatother
\usepackage{xcolor}
\newcommand{\reviewertwo}[1]{{\color{black}#1}}
\newcommand{\reviewerone}[1]{{\color{black}#1}}

\usepackage[hidelinks]{hyperref}

\usepackage{authblk}

\usepackage{graphicx}
\usepackage{babel}
\usepackage{amsmath}

\usepackage{amssymb}
\usepackage{microtype}
\usepackage{physics}


\usepackage{geometry}
\raggedright

\usepackage[T1]{fontenc}

\usepackage{libertine}

\usepackage{setspace}
\usepackage[]{footmisc}

\usepackage[font=normalsize,font=singlespacing,labelfont=bf,justification=raggedright]{caption}

\usepackage{titlesec}
\titleformat*{\section}{\bf}
\titleformat*{\subsection}{\it}
\titleformat*{\subsubsection}{\it}

\begin{document}

\title{The Time in Thermal Time}
\author{Eugene Y. S. Chua\thanks{Division of the Humanities and Social Sciences, California Institute of Technology, Pasadena, CA, USA}$\,$\thanks{School of Humanities, Nanyang Technological University, Singapore} \\ \small{\textit{eugene.chua@cantab.net}} \\ 

\bigskip 

\small{\textit{Accepted for publication in the Journal for General Philosophy of Science}, \\ \textit{as part of the Special Issue: On Time in the Foundations of Physics} \\ \textit{(eds. Andrea Oldofredi \& Cristian Lopez)} \\ \bigskip Please cite final version when available.}}

% add this for a custom date.
% if commented out, today's date will be added.
% if inside is blank, then no date will be shown i.e. \date{}
%\date{Month DD, YYYY}

\maketitle

\thispagestyle{empty}

\begin{abstract}
    \noindent \reviewertwo{Preparing general relativity for quantization in the Hamiltonian approach} leads to the `problem of time,' rendering the world fundamentally timeless. One proposed solution is the `thermal time hypothesis,' which defines time in terms of states representing systems in thermal equilibrium. On this view, time is supposed to emerge thermodynamically even in a fundamentally timeless context. Here, I develop the worry that the thermal time hypothesis requires dynamics -- and hence time -- to get off the ground, thereby running into worries of circularity.
\end{abstract}

\section{Preamble: the problem of time}\label{1}

In canonical quantum gravity, we formulate general relativity in Hamiltonian form with the appropriate constraints and quantize -- just as we obtained quantum mechanics from classical mechanics in the context of non-gravitational physics.\footnote{There is a literature surrounding whether this is the right approach to quantization (e.g. Pitts (2014), Pooley \& Wallace (2022)); I won't pursue this discussion here and will assume the standard Dirac quantization approach is correct.} However, dramatically, the resulting Wheeler-DeWitt equation appears absent of dynamical evolution. Schematically:
\begin{equation} \label{eq:1}
    \hat{H}|\Psi\rangle = 0
\end{equation}
where $\hat{H}$ is the Hamiltonian, and $|\Psi\rangle$ is associated with the wave-functional $\Psi$ representing both matter content and geometry. More concretely, we find an infinite set of Hamiltonian constraints of the form:
\begin{equation}\label{hamcon}
    \hat{H}(f) = \int d^3x \hat{H}(x) f(x)
\end{equation}
where $\hat{H}(x)$ is the Hamiltonian constraint density at each point of space, and $f(x)$ is a test function. \textit{All} these constraints must vanish, i.e. something like \eqref{eq:1} holds for all these constraints so that \textit{any} physical Hamiltonian vanishes.\footnote{Thanks to an anonymous reviewer for suggesting this point.}

Hence, while the usual Schrödinger equation
\begin{equation} \label{eq:2}
    \hat{H}|\Psi\rangle = i\hbar \frac{\partial |\Psi\rangle}{\partial t}
\end{equation}
describes a wave-function $|\Psi\rangle$'s time-evolution under some physical Hamiltonian, the Wheeler-DeWitt equation -- its quantum gravity analogue -- does not. Assuming this equation describes the fundamental state of
affairs, the fundamental ontology contains no reference to time. Yet, our familiar physical systems are manifestly evolving \textit{in time}. This is the problem of recovering time-evolution from fundamentally timeless ontology: the \textit{problem of time}.\footnote{See Kuchar (1991), Isham (1993), Kuchar (2011), Anderson (2017), or Thébault (2021).} 

There have been many attempts to solve, resolve, or dissolve the problem. Here, I'll assess one: the \textit{thermal time hypothesis} (\textbf{TTH}) from Connes \& Rovelli (1994). In their words:
\begin{quote}
   A radical solution to this problem [...of this absence of a fundamental physical time at the ... generally covariant level\footnote{\reviewertwo{The generally covariant level here refers to the general relativistic, or gravitational, regime.}}...] is based on the idea that one can extend the notion of time flow to generally covariant theories, but this flow depends on the thermal state of the system ... provided that we:
    \begin{enumerate}
        \item interpret the time flow as a one-parameter group of automorphisms of the observable algebra... 
        \item ascribe the temporal properties of the flow to thermodynamical causes, and therefore we \textit{tie the definition of time to thermodynamics} and...
        \item take seriously the idea that in a generally covariant context \textit{the notion of time is not state-independent, as in non-relativistic physics, but rather depends on the state in which the system is in}. (Connes \& Rovelli 1994, 2901, emphasis mine)
    \end{enumerate}    
 \end{quote}
I take these to be core tenets of \textbf{TTH}: despite the problem of time and timeless context, time emerges due to thermodynamic origins. Starting with a world without time in its fundamental ontology, if we can find systems in special states within the fundamentally timeless ontology -- \textit{Kubo-Martin-Schwinger (KMS) thermal states} -- then time's emergence comes in three parts: first, we use these states to define a privileged one-parameter automorphism group given a certain algebraic structure, second, we interpret this parameter as a bona fide time parameter, and third, we explain this choice of interpretation in terms of thermodynamic considerations.\footnote{Rovelli (personal correspondence) distances himself from this stronger hypothesis, suggesting that it merely provides analysis of `timely' notions associated with thermodynamics, but \textit{doesn't define} time using thermodynamics. Importantly, he claims that \textbf{TTH} already assumes that some notion of time (perhaps relational clocks) is definable. This weaker TTH avoids the conceptual problems I'll raise. However, it's contrary to a natural reading of the above passage. I'll set this weaker \textbf{TTH} aside -- it already assumes some (presently absent) resolution of the problem of time by assuming that time can be defined.}

To my knowledge, philosophers have not dedicated much attention to \textbf{TTH} beyond Swanson (2014 ch. 5, and 2021). Furthermore, while Swanson (2021) focuses largely on technical issues with \textbf{TTH}, I want to emphasize a core conceptual issue. I thus aim to complement Swanson's discussion and generate more interest in \textbf{TTH} as a solution to the problem of time, and more generally, the nature of the problem of time itself.

Prima facie, \textbf{TTH} elegantly resolves the problem of time. It's commonly accepted that systems are in thermal equilibrium if certain thermodynamic parameters are unchanging over time, but \textbf{TTH} reverses this observation by proposing that the time parameter is \textit{defined} \textit{by} systems in states of thermal equilibrium (thermal states), which \textbf{TTH} understand to be states satisfying the KMS condition. Furthermore, \textit{every} state of interest is supposed to uniquely pick out some such parameter. As long as we find systems in thermal states in a timeless world, we can recover (thermal) time. 

\reviewertwo{\textbf{TTH}'s strategy for time's emergence can be understood along the lines of Lam \& W\"uthrich (2018, 2021)'s proposal of \textit{functional realization} or functionalism.\footnote{\reviewertwo{See Huggett (2021) and Salimkhani (2023) and references therein for a more synoptic view of the literature on spacetime emergence. See also Baron, Miller, and Tallant (2022, Ch. 6) for critical assessment of such approaches.}} On this view, something at the higher level emerges from something at the lower level given two general steps:
\begin{enumerate}
    \item The higher-level properties or entities are functionalized, that is, given a definition in terms of their functional role. 
    \item An explanation is provided for how lower-level properties and entities can fill this functional role. 
\end{enumerate}
In the case of \textbf{TTH}, the lower level consists of states satisfying a formal KMS condition and their associated one-parameter modular automorphism groups. The higher level includes the property of time, understood functionally (and minimally) as that which plays a dynamical role in parametrizing time-evolution generated by the action of physical Hamiltonians. If \textbf{TTH} can then explain how the lower-level modular group in fact fills the higher-level dynamical role played by time, without appealing to notions to do with time or dynamics, then we can say that the \textbf{TTH} succeeds in addressing the problem of time. \textbf{TTH}'s main explanatory strategy -- and why \textbf{TTH} involves \textit{thermal} time -- is to argue that states satisfying the formal KMS condition \textit{are} states representing systems genuinely in thermal equilibrium, with which we then explain how the modular group can play the dynamical role of time. As we'll see, what's needed more generally is simply a reason -- thermodynamic or not -- for adopting a dynamical interpretation of the modular group, or, equivalently, why the associated modular `Hamiltonian' (which we'll see later) at the lower level should be interpreted as a generator of the usual time-translations at the higher level.} 

Overall, I'll argue that \textbf{TTH}'s strategy is unsatisfactory as a solution to the problem of time: it's either circular or remains yet unjustified in securing time's emergence. The general worry is this: a dynamical interpretation for the modular group associated with KMS states, which would let the modular group fill the dynamical role played by time at the higher level, is generally not guaranteed. However, I'll argue that pinning down such an interpretation requires background considerations about dynamics, and hence time, thereby failing the explanatory desideratum necessary for securing time's emergence.

I'll press my concern through three perspectives with which one might try to pin down such a dynamical interpretation: classical thermodynamics, Rovelli's (1993) proposed `intrinsic' thermodynamics, and considerations about the algebraic structure itself. From the first perspective, equilibrium is intrinsically tied up with the concept of time, which runs quickly into circularity should we attempt to use the equilibrium concept to define time. The second perspective seeks to remedy this by defining equilibrium in terms of concepts not obviously tied up with time. I'll argue that this proposed definition -- of `intrinsic' equilibrium -- is classically motivated by time-scale and dynamical considerations; furthermore, free from such considerations, the definition is neither necessary nor sufficient for thermal equilibrium. Finally, I'll argue that the modular group itself does not necessitate -- and in fact, does not generally have -- a dynamical interpretation. In the cases where it does, it is again justified by background considerations which appeal to time, dynamics, or spacetime. 

In short, it seems that \textbf{TTH} cannot succeed yet without employing concepts to do with time, thus interpreted. If we already have time in the background, though, then we run into circularity. For now, either \textbf{TTH} does not take off, or it's assuming the very thing it intended to deliver and is hence redundant. \reviewertwo{I conclude with some brief comments on how \textbf{TTH} might attempt to circumvent this worry in order to satisfy the explanatory desideratum.}

\section{The time from thermal time}\label{2}

\subsection{The Heisenberg picture}\label{2.1}

While not necessary, \textbf{TTH} finds its most natural setting when viewed through the Heisenberg picture of quantum mechanics because of its focus on an operator-algebraic approach.\footnote{The C$^*$ algebraic approach does not itself depend on any particular picture of quantum mechanics, though Connes \& Rovelli (1994, 2901) emphasizes that \textbf{TTH} is to be interpreted in terms of a generalized Heisenberg picture.} In standard physics, the Heisenberg picture takes states $|\Psi\rangle$ in Hilbert space $\mathcal{H}$ to be time-\textit{in}dependent. What evolves unitarily over time -- and are of interest -- are the time-\textit{dependent} observables $\mathcal{O}$ -- representing possible measurement outcomes of physical quantities one might make on systems -- which are represented by self-adjoint linear operators $A \in \mathcal{A}$ acting on $\mathcal{H}$.\footnote{More specifically, $\mathcal{A}$ has the structure of $\mathcal{B}(\mathcal{H})$, that is, the algebraic structure of bounded linear operators acting on $\mathcal{H}$.} However, the physical interpretation of $\mathcal{A}$ is similar to those in the Schr\"odinger picture: they are the possible physical quantities attributable to systems at a time (or at some spacetime region).\footnote{Rovelli \& Smerlak (2011, 6) suggests that ``in quantum gravity the pure states can be given by the solutions of the Wheeler-DeWitt equation, and observables by self-adjoint operators on a Hilbert space defined by these solutions.'' That is, the observables of interest in quantum gravity might be a global algebra of observables across spacetime. Crucially, such Hilbert spaces (and global algebras) remain elusive.} 

$\mathcal{A}$ evolves unitarily according to the Heisenberg equation of motion: 
\begin{equation}\label{eq:4}
    \frac{\partial A }{\partial t} = \frac{i}{\hbar}[H, A]
\end{equation}
and: 
\begin{equation}\label{eq:5}
    A (t) = U^\dag(t) \, A (0) \,U(t)
\end{equation}
\noindent where $U(t) = e^{-iHt/\hbar}$ is the unitary operator, $[\cdot , \cdot]$ is the commutator defined as $[A, B] = AB - BA$, and $H$ is the Hamiltonian. 

\subsection{C$^*$ algebra: from abstract to concrete}\label{2.2}

The Heisenberg picture lends naturally to an \textit{algebraic} interpretation of quantum mechanics. The structure of how $\mathcal{A}$ acts on $\mathcal{H}$ can be understood using abstract algebra: it's a non-commutative C$^*$ algebra, specifically, a von Neumann algebra.\footnote{For technical exposition, see Bratteli \& Robinson (1987 and 1997). For exposition targeted at philosophical audiences, see Ruetsche (2011a, Ch. 4).} Furthermore, this abstract algebraic structure of C$^*$ algebras can be represented in terms of corresponding \textit{concrete} C$^*$ algebras via the familiar Hilbert space structure. 

Given an abstract C$^*$ algebra $\mathcal{C}$,\footnote{An \textit{abstract} C$^*$ algebra is a set $\mathcal{C}$ of elements -- such as the observables we're interested in -- which satisfies various formal properties: it's closed under addition, scalar multiplication, (non-commutative) operator multiplication, involution operation $^*$, and equipped with a norm $\vert\vert \; \vert\vert$ satisfying $\vert\vert A^*A \vert\vert = \vert\vert A \vert\vert^2$ and $\vert\vert AB \vert\vert \leq \vert\vert A \vert\vert \: \vert\vert B \vert\vert$ for all $A, B \in \mathcal{C}$.} one can define \textit{states} $\omega$ over $\mathcal{C}$: positive normalized linear functionals such that
\begin{equation}
    \omega: \mathcal{C} \to \mathbb{C}
\end{equation}
Since it's normalized, positive, linear, and real-valued on self-adjoint elements, a natural interpretation of $\omega(\mathcal{C})$ is to understand it as assigning \textit{expectation values} to the physical quantities $\mathcal{A}$ -- the elements of $\mathcal{C}$ -- in state $\omega$.

Now we connect these abstract notions to concrete physics: this means finding counterparts in $\mathcal{H}$ to the algebraic structure and states via  \textit{representations}. Notably, for each $\omega$ on $\mathcal{C}$, the Gelfand-Naimark-Segal (GNS) construction provides a representation $\pi_\omega(\mathcal{C})$ of $\mathcal{C}$ in some Hilbert space $\mathcal{H_\omega}$.\footnote{A \textit{representation} is a $^*$-homomorphism $\pi: \mathcal{D} \to \mathcal{B}(\mathcal{H})$ where $\mathcal{D}$ is some abstract algebra, and $\mathcal{B}(\mathcal{H})$ is the set of bounded linear operators on $\mathcal{H}$.} Within $\mathcal{H}_\omega$, we find a cyclic and separating vector $|\Psi_\omega\rangle \in \mathcal{H}_\omega$ such that:\footnote{A vector $|\Psi\rangle \in \mathcal{H}$ is \textit{cyclic} for $\mathcal{C}$ just in case $\mathcal{W}|\Psi\rangle$ is dense in $\mathcal{H}$. $|\Psi\rangle \in \mathcal{H}$ is said to be \textit{separating} for $\mathcal{C}$ just in case $A|\Psi\rangle = 0$ implies $A = 0$ for any $A \in \mathcal{C}$.}
\begin{equation}
    \omega(A) = \langle \Psi_\omega | \pi(A) | \Psi_\omega \rangle 
\end{equation}
In Ruetsche's words, ``the expectation value the state $\omega$ assigns the algebraic element $A$ is duplicated by the expectation value the vector $|\Psi_\omega\rangle$ assigns to the Hilbert space operator $\pi(A)$". (2011a, 92) This shows that every abstract $\omega$ has concrete counterparts as vectors in some Hilbert space given by the GNS representation. \reviewerone{However, $|\Psi_\omega\rangle$ is separating if and only if $\pi_\omega$ is \textit{faithful}. Thankfully, one can always find states guaranteeing a faithful representation; these are faithful states.}\footnote{I assume that this restriction to faithful states is unproblematic, following Connes \& Rovelli (1994). See Feintzeig (2023, $\S2.3$) for discussion. Swanson (2021, 286) notes that nontrivial C$^*$ algebras have no \textit{pure} faithful states, and worries that \textbf{TTH} might thereby require some ignorance interpretation of mixed states. However, Wallace (2012) argues that mixed states needn't demand such interpretations. See also Chen's (2021) and Chua \& Chen's (ms) discussion of density matrix realism, which is a realist picture of quantum mechanics employing mixed states without assuming that such mixed states represent ignorance.} Roughly, faithful representations preserve abstract algebraic structure in concrete settings.\footnote{More specifically, faithful states ensure $\pi$ is a $^*-$homomorphism to a subset of bounded operators on $\mathcal{H}$, i.e. that $\pi$ is a faithful representation. Faithful states satisfy the condition that $\omega(A^*A) = 0$ entails $A = 0$ for all $A \in \mathcal{C}$.} 

These results establish correspondence between the abstract states and algebraic structure of $\mathcal{C}$ and concrete representations in terms of bounded linear operators $A$ acting on $\mathcal{H}$, i.e. $\mathcal{B}(\mathcal{H})$. In other words, the existence of faithful representations (via the GNS construction) associated with faithful states guarantees that any abstract C$^*$-algebra is isomorphic to a concrete C$^*$-algebra. For the remainder of this paper, I'll furthermore focus on concrete \textit{von Neumann} algebras $\mathcal{W}$ per Connes \& Rovelli,\footnote{These are concrete C$^*$ algebras closed under the weak operator topology satisfying $\mathcal{W} = \mathcal{W}''$. For an algebra $\mathcal{D}$ of bounded operators on $\mathcal{H}$, its commutant $\mathcal{D}'$ is the set of all bounded operators on $\mathcal{H}$ commuting with every element of $\mathcal{D}$. If $\mathcal{D}$ is an algebra, so is $\mathcal{D}'$. $\mathcal{D}''$ is the \textit{double commutant}, the set of all bounded operators in $\mathcal{D}'$ commuting with $\mathcal{D}'$.} which allows us to use the tools from modular theory (which depend on the structure of $\mathcal{W}$). 

Connes \& Rovelli also restricts attention to \textit{normal states}: states on some given $\mathcal{W}$ which also satisfy countable additivity. Normal states are represented as density operators $\rho$ with $Tr(\rho) = 1$ in $\mathcal{H}$:
\begin{equation}
    \omega(A) = Tr(\rho A) 
\end{equation}
for all $A \in \mathcal{W}$. This restriction is likely motivated by a demand that the algebraic formalism be given clear physical meaning.\footnote{See Ruetsche (2011b).} After all, (7) recovers the standard way of deriving expectation values -- observed statistics -- of measurements associated with observables $\mathcal{A}$:
\begin{equation}
    \langle A \rangle = Tr(\rho A) 
\end{equation}

\subsection{From kinematics to possible dynamics}\label{2.3}

So far we've focused on algebraically representing the kinematics of quantum theory -- the structure of $\mathcal{A}$ acting on $\mathcal{H}$ and their expectation values -- \textit{at a time} (or spacetime region). But $\mathcal{W}$ also provides something that \textit{could} be understood as time-evolution, via modular theory. Crucially, \textit{any faithful, normal $\omega$} defines a \textit{unique} one-parameter group of automorphisms of $\mathcal{W}$:
\begin{equation}
    \alpha_t^\omega: \mathcal{W} \to \mathcal{W}
\end{equation} 
for real $t$. Given a concrete $\mathcal{W}$ defined by faithful, normal $\omega$ via the GNS construction, the Tomita-Takesaki theory provides a unique $\alpha_t$ in terms of two modular invariants generated from the adjoint conjugation operation $^*$. The theory guarantees the existence of a well-defined operator $S$:
\begin{equation}
    SA |\Psi\rangle = A^* |\Psi\rangle
\end{equation}
and that $S$ has a unique polar decomposition:\footnote{See Takesaki (1970).}
\begin{equation}
    S = J\Delta^{1/2}
\end{equation}
where $J$ is antiunitary and $\Delta$ is a self-adjoint positive operator. $\alpha_t$, associated with the defining state $\omega$, is defined by: 
\begin{equation}\label{eq:6}
    \alpha_t^\omega A = \Delta^{it} A \Delta^{-it}
\end{equation}
and this uniquely defines a strongly continuous one-parameter unitary group of automorphisms on $\mathcal{W}$, parametrized by $t \in \mathbb{R}$, which is also called the \textit{modular group}. Associated with the group is a modular `Hamiltonian' $\text{log} \; \Delta$ which is the generator of the modular group.\footnote{One can see this via Stone's theorem, which states that every strongly continuous one-parameter unitary group, $\Lambda(t)$, is associated with a corresponding (possibly unbounded) generator via $e^{iHt}$. If $t$ is interpreted as time, then $H$ is the Hamiltonian.} The modular `Hamiltonian' is introduced in scare-quotes to distinguish it from the physical Hamiltonian which is the generator of the time-translation group, that is, the driver of time-evolution via e.g. the Heisenberg equation \eqref{eq:5}. In general, the two notions \textit{cannot} be identified, in which case the associated modular group is \textit{not} the time-translation group, but some other arbitrary unitary group. As Schroer puts it, this modular `Hamiltonian' ``is always available in the mathematical sense but allows a physical interpretation only in those rare cases when it coincides with one of the global spacetime generators" (2010a, 114) and that ``modular Hamiltonians give rarely rise to geometric movements (diffeomorphisms)" (2010b, 306). In other words, the modular `Hamiltonian' does not necessarily have a clear dynamical role to play -- qua physical Hamiltonian driving time-evolution -- outside of certain special circumstances.

Crucially, this means that $\alpha_t^\omega$ \textit{can} (but need not always!) be given dynamical meaning because $\Delta^{-it}$ \textit{can} (but need not always!) be interpreted as time-translation operators. In other words, $t$ isn't always time. If the physical Hamiltonian -- associated with the usual time-evolution via \eqref{eq:5} -- can be \textit{identified} with the modular `Hamiltonian', then \eqref{eq:6} is equivalent to \eqref{eq:5}. This, of course, is \textit{not} guaranteed: not all unitary operators implement time-translation. As Ruetsche (2011) puts it, ``every faithful normal state on a von Neumann algebra satisfies the modular/KMS condition with respect to exactly one flow, although not necessarily one naturally read as a group of time translations." (164) The crucial step, in interpreting the modular group defined by any faithful, normal state dynamically, is to interpret its parameter $t$ as playing \textit{the same role as physical time in time-evolution} in the usual Heisenberg equation. \reviewertwo{Put another way, and returning briefly to the aforementioned explanatory desideratum needed to secure functional realization and time's emergence, what we need is an explanation for why the $t$ associated with $\alpha_t^\omega$ actually succeeds in playing the dynamical role that physical time plays in time-evolution.} (More on this in $\S\ref{3.2}$.)

Furthermore, even if we \textit{can} interpret $\alpha_t^\omega$ dynamically in the sense above, these are very \textit{special} dynamics: any faithful, normal state $\omega$ is invariant under the `flow' of $\alpha_t^\omega$:
\begin{equation}
\omega(\alpha_t^\omega A) = \omega(A)
\end{equation}
Interpreted dynamically, $\alpha_t^\omega$ leaves $\omega$ unchanged over time. But systems in arbitrary states need \textit{not} be unchanging over time; in these cases, the dynamics associated with $\alpha_t^\omega$ doesn't describe the dynamics of that system. Put simply: the two notions of dynamics -- the `dynamics' of $\alpha_t^\omega$ and the system's actual dynamics -- don't always `align' and we're \textit{not} always justified to interpret $\alpha_t^\omega$ as the system's actual dynamics. (More on this in $\S\ref{3.1}$ and $\S\ref{3.2}$.)

\subsection{Justifying a dynamical interpretation: from thermal states to KMS states}\label{2.4}

Importantly, there is one clear case in ordinary physics when we \textit{are} physically justified in interpreting $\alpha_t^\omega$ dynamically: when systems are in \textit{thermal states}. In such cases, systems are time-translation-invariant (i.e. stationary) and possess certain thermodynamic properties e.g. being at constant temperature. In standard physics, these notions are defined via background time and dynamics. \textit{Given} the special kind of dynamics associated with such systems -- dynamics which doesn't change the system's thermodynamic state over time -- the associated modular group automorphisms $\alpha^\omega_t$ can \textit{then} be interpreted as the actual dynamics for systems in such a state. For this special case, the dynamics associated with $\alpha^\omega_t$ seems to `align' with the dynamics of systems in thermal equilibrium.

When do we know that states $\omega$ are thermal? It turns out that one can understand thermal states (with inverse temperature $\beta$, $0 < \beta < \infty$) as states satisfying a formal KMS condition, i.e. KMS states.\footnote{$\beta = \frac{1}{k_bT}$, where $T$ is the system's temperature, and $k_b$ is Boltzmann's constant.} KMS states satisfy the following conditions: for any $A, B \in \mathcal{W}$, there exists a complex function $ F_{A,B}(\textit{z})$, analytic in the strip $\{ z \in \mathbb{C} \; | \; 0 < \text{Im } \textit{z} < \beta \}$ and continuous on the boundary of the strip, such that for all $t \in \mathbb{R}$:
\begin{equation}
    F_{A, B}(t) = \omega(\alpha_t^\omega(A)B)
\end{equation}
\begin{equation}
    F_{A, B}(t + i\beta) = \omega(B\alpha_{t}^\omega(A))
\end{equation}
\begin{equation}\label{eq:17}
    \omega(\alpha_t^\omega(A)B) = \omega(B\alpha_{t}^\omega(A))
\end{equation}
The KMS condition is arcane, but a physical anchor -- and one big reason (though not the only reason) \textit{why} we can interpret KMS states as thermal states -- is the fact that KMS states are formally equivalent, in the finite-dimensional case, to Gibbs states $\rho_\beta$. This is the quantum generalization of statistical states for systems in thermal equilibrium at constant inverse temperature $\beta$ with a physical Hamiltonian $H$: 
\begin{equation}\label{eq:18}
    \rho_\beta = \frac{e^{-\beta H}}{Tr(e^{-\beta H})}
\end{equation}
For any operator $A \in \mathcal{A}$, the expectation value for that observable for this system is:
\begin{equation}\label{eq:19}
    \langle A \rangle_\rho = Tr(\rho_\beta A) = \frac{Tr(e^{-\beta H}A)}{Tr(e^{-\beta H})}
\end{equation}
$\rho_\beta$ satisfies the KMS condition. Interpreting $\omega$ in terms of $\rho_\beta$ via \eqref{eq:19}, and $\alpha_t^\omega(A)$ as $e^{iHt}Ae^{-iHt}$ per \eqref{eq:5}, we get, for operators $A, B \in \mathcal{W}$: 
\begin{equation}\label{eq:20}\begin{split}
    Z^{-1}Tr({e^{-\beta H}} e^{iHt} A e^{-iHt}B) = Z^{-1}Tr(e^{-\beta  H}B{e^{iH(t+i\beta)}Ae^{-iH(t+i\beta)}}) 
\end{split}
\end{equation}
where $Z = Tr(e^{-\beta H})$ is the partition function of $\rho_\beta$.\footnote{This uses the fact that the Hamiltonian commutes with itself, and that the trace is cyclic.} From \eqref{eq:20} we see that $\rho_\beta$ satisfies the KMS condition \eqref{eq:17}. Note, again, that we must first justify interpreting $\alpha_t^\omega$ dynamically as the time-translation group associated with the Heisenberg equation, and not just any arbitrary unitary group. In this case, since we started with a physical Hamiltonian which plays the dynamical role in the Heisenberg equation, e.g. by encoding the equations of motion, we have a physical reason for why $\alpha_t^\omega$ can naturally be interpreted as a time-evolution operator. 

For finite-dimensional quantum systems, $\rho_\beta$ uniquely describes systems satisfying the KMS condition.\footnote{See Emch \& Liu (2002, 351--352). In infinite-dimensional quantum systems, the trace is ill-defined, and so $\rho_\beta$ is likewise ill-defined. Crucially, the KMS condition can still hold.} This imbues the purely syntactic KMS condition (as Emch \& Liu (2002, 351) describes it) with physical meaning, motivating the \textit{physical equivalence} of KMS states and thermal states: the $t$ in $\alpha_t^\omega$ can be interpreted as the time along which systems stay in thermal equilibrium. The interpretation of KMS states as thermal states is further motivated by the discovery that KMS states also satisfy various stability and passivity conditions we typically associate with thermal states, and that these properties hold even in the infinite-dimensional case (where $\rho_\beta$ is ill-defined).\footnote{$\omega$ is invariant under the flow of $\alpha^\omega_t$; interpreted as dynamical flow, it captures the idea that equilibrium states are stationary and don't change over time. Other examples: such states don't change in free energy over time (thermodynamic stability), remain (over time) in arbitrarily close stationary states under small perturbations (dynamical stability), and are passive under any finitely long local perturbations of its dynamics (passivity). See Emch \& Liu (2002, 355).} 

Importantly, \textit{any} $\omega$ satisfies the KMS condition relative to the modular group defined by itself, i.e. $\alpha^\omega_t$,\footnote{See Bratteli \& Robinson (1997).} for $\beta = 1$.\footnote{Ruetsche (2011a, Ch. 7, fn. 23) notes that states satisfying the KMS condition for $\beta = 1$ also satisfy it for arbitrary $\beta > 0$.} Na\"ively, this seems to overgeneralize: \textit{any} state \textit{is} a KMS state, even the state of my cooling coffee. However, the appropriate statement is that any state \textit{can} be a KMS state: there are \textit{some possible} dynamics which keeps a system in thermal equilibrium. This is just to reiterate that $\alpha^\omega_t$ can but needn't necessarily be interpreted dynamically. It needn't align with a system's actual dynamics. Furthermore, when they don't align, $\omega$ will not be a KMS state with respect to the actual dynamics (i.e. it's not really in thermal equilibrium). (I'll elaborate in $\S\ref{3.1}$.)

Returning to when we can interpret $\alpha_t^\omega$ dynamically, it seems that we're justified to do so when we're justified to interpret a system as being in thermal equilibrium. \textit{If} we know that the system's dynamics -- associated with thermal equilibrium -- `aligns' with the special dynamics described by $\alpha_t^\omega$, or whether the physical Hamiltonian can be identified with the modular `Hamiltonian', \textit{then} we can interpret $\alpha_t^\omega$ dynamically. This is the case only in very special cases, for instance, when we consider immortal uniformly accelerating observers restricted to the right wedge of Rindler spacetime, where the vacuum state looks like a thermal state due to the Unruh effect. There, the modular `dynamics' $\alpha_t^\omega$, and the associated modular `Hamiltonian', can be identified with the natural dynamics of such an observer in terms of Rindler wedge-preserving Lorentz boosts (and the associated physical Hamiltonian).\footnote{See Earman (2011) or Swanson (2021) for discussion.}

\subsection{The thermal time hypothesis}\label{2.5}

So far I've introduced everything in a standard quantum mechanical context, where there is some assumed background time (or spacetime). In the timeless context, however, there's no time with which we may determine systems to be in thermal equilibrium, and no straightforward way to understand the physical meaning of KMS states (and hence interpret the associated $\alpha^\omega_t$ dynamically). 

\textbf{TTH} reverses this situation. Instead of defining thermal equilibrium and KMS states \textit{in terms of time}, Connes \& Rovelli hypothesizes that we \textit{define} the modular group parameter to \textit{be} time and \textit{justify} this interpretation via reasoning to do with thermal equilibrium. Notably, this implicitly assumes the applicability of the C$^*$ algebraic structure even in the timeless context.

\textbf{TTH} is motivated by the aforementioned fact that any faithful, normal, state $\omega$ defines a preferred one-parameter group of automorphisms $\alpha^\omega_t$. \textit{If} we're further justified in interpreting $\omega$ as a thermal state $\rho_\beta$, \textit{then} we can interpret the dynamics of $\alpha^\omega_t$ as being generated by a `thermal' physical Hamiltonian $H = -ln \rho_\beta$, a Hamiltonian that describes the dynamics of a system in thermal equilibrium.\footnote{For more details, see Paetz (2010, $\S$4.2 and $\S$5.2).} Crucially, in this case, the physical Hamiltonian is defined \textit{in terms of $\rho_\beta$}, and is definitionally equivalent to the modular `Hamiltonian'. This is contrary to the usual understanding where the physical Hamiltonian is defined and interpreted \textit{prior} to $\rho_\beta$, and is used to \textit{define} $\rho_\beta$. Given genuine thermal states, this move is unproblematic and a physical Hamiltonian can indeed be extracted in non-generally covariant contexts. The further, and more daring, claim of \textbf{TTH} is that we can do \textit{the same thing} in generally covariant contexts: start with formal KMS states, interpret them even in the timeless setting to be genuine thermal states, then define a physical Hamiltonian and the associated time-translation group in terms of the modular `Hamiltonian' and modular group associated with such states.

To sum up \textbf{TTH}: in the generally covariant context of quantum gravity, where the problem of time looms, we appeal to the C$^*$ algebraic structure and hypothesize that \textit{the flow of time is defined by the unique one-parameter state-dependent modular automorphism group} $\alpha^\omega_t$; dynamical equations can be defined in terms of this flow (e.g. via the `thermal' Hamiltonian above). Systems in thermal equilibrium thus define time even in the timeless setting, providing a path towards tackling the problem of time.

\section{The time in thermal time}\label{3}

While the foregoing technical details are daunting, the conceptual point is simple: in standard quantum mechanics and classical thermodynamics, we always have some background (space-)time, with which we can define dynamical notions such as equilibrium, time-evolution, stationarity, etc. In the generally covariant setting we don't have such a time parameter. But, \textit{if} we had access to the structure of $\mathcal{W}$, then any faithful, normal state $\omega$ over $\mathcal{W}$ defines $\alpha^\omega_t$ according to which it's a KMS state, which comes with an associated unitary group, and an associated `Hamiltonian'. Connes \& Rovelli's proposal is that we \textit{first} interpret these states $\omega$ as thermal equilibrium states, \textit{then} interpret their dynamics $\alpha_t^\omega$ as thermal equilibrium dynamics i.e. the parameter $t$ as time. Time is defined \textit{in terms of} thermal equilibrium via $\alpha_t^\omega$. 

Swanson (2021) has already pointed out some technical challenges for this program.\footnote{See also Swanson (2014), Paetz (2010, Ch. 7).} Here, I emphasize a further \textit{conceptual} challenge. Essentially, \textbf{TTH} tries to define time \textit{in terms of} the modular group of KMS states by interpreting KMS states as states of genuine thermal equilibrium, by working with the C$^*$ algebraic structure. To avoid circularity, and to genuinely tackle the problem of time, the justification for this interpretation had better not require background temporal notions to be physically meaningful to begin with. Otherwise, they only define time insofar as time has already been defined -- a circularity \textit{par excellence}. However, I'll argue that three different perspectives towards the \textbf{TTH} all point to this conceptual worry.

Let me first sketch the core of the worry. For Connes \& Rovelli, ``an equilibrium state is a state whose modular automorphism group is the time translation group" for the non-generally covariant context (1994, 2909), and \textbf{TTH} asserts that this carries over to the generally covariant context. However, when are we allowed to interpret the modular group as the \textit{time}-translation group, as opposed to any other unitary group (e.g. those associated with spatial translations)? Put in terms of the explanatory desideratum sketched in \S1, why are we justified in thinking that the modular group fills the dynamical role played by time at the higher level?

As I've already emphasized, $\alpha^{\omega}_t$ cannot be interpreted dynamically automatically. Furthermore, even if it \textit{is} interpreted dynamically, it is a very special sort of dynamics for the associated $\omega$. Earman \& Ruetsche echo this concern: ``the modular group determined by an arbitrary faithful normal state on a von Neumann algebra may lack a natural dynamical interpretation, in which case scare quotes should be understood when referring to $\beta$ as the inverse temperature." (2005, 570) That is, we're not entitled to interpret \textit{any} (faithful normal) state satisfying the KMS condition as being in thermal equilibrium, or having (equilibrium) thermodynamic properties, without further justification. Even in non-generally covariant contexts, $\alpha^{\omega}_t$ might not align with a system's actual dynamics, and hence might not be interpretable dynamically. We need some further \textit{physical argument} for why systems in some arbitrary state ought to be interpreted as having the dynamics associated with $\alpha^{\omega}_t$ -- for why its dynamics `aligns' with $\alpha^{\omega}_t$'s dynamics.

This point was already emphasized by Haag et al (1967), who first connected the KMS condition to thermodynamic equilibrium: 
\begin{quote}
    We assumed the existence of an automorphism $A \to A_t$ for which $\omega(A)$ is invariant. It then follows that there exists a unitary operator $U(t) = e^{-iHt/\hbar}$ on $\mathcal{H}$, which implements this automorphism. \textit{This does not mean, however, that the system actually moves according to this automorphism.} It only means that it's possible to choose the dynamics, i.e. the interparticle forces and the external forces, such that with these forces the system in the state $\omega(A)$ would be in equilibrium. If the forces happen to be different, the automorphism $A \to A_t$ is not a time translation, $H$ is not the Hamiltonian of the system and the state $\omega(A)$ is not stationary. (1967, 235, emphasis mine)
\end{quote}
Put another way, we're justified in taking $\alpha^{\omega}_t$ seriously as dynamics only when we already have some \textit{prior} determination that the system is \textit{already} in thermal equilibrium. Likewise, Swanson (2021, 12) points out:
\begin{quote}
    Any statistical state determines thermal dynamics according to which it is a KMS state, however, if $\rho$ is a non-equilibrium state, the resultant thermal time flow does not align with our ordinary conception of time. By the lights of thermal time, a cube of ice in a cup of hot coffee is an invariant equilibrium state! The same problem arises in the quantum domain -- only for states which are true equilibrium states will the thermal time correspond to physical time.
\end{quote}
In other words, for \textbf{TTH} to take off, it must rule out the fact that any arbitrary faithful, normal state defines \textit{some} `thermal time'. It must restrict \textbf{TTH} only to the physically meaningful thermal times defined by a privileged class of states over $\mathcal{W}$, which are `really' equilibrium states. After all, as I've emphasized, genuine thermal states are the only states for which $\alpha_t^\omega$ aligns with a system's actual dynamics. But which states are `really' equilibrium states? And can an explanation for why they are `really' equilibrium states be given without appealing to time to begin with?

\subsection{The time in standard thermodynamic accounts of equilibrium}\label{3.1}

Let me start, briefly, with the standard account of thermodynamic equilibrium. The usual way (in standard quantum mechanical contexts) of picking out equilibrium states refer to thermodynamic properties such as stationarity, stability, and passivity. Emch \& Liu (2002, 355) observes that these properties typically requires that the state ``is assumed tacitly to be stationary with respect to a specified dynamics $\alpha$".\footnote{See fn. 20 for some of these conditions.} In other words, bona fide equilibrium states appear to \textit{be defined implicitly in terms of some background time parameter}. 

More generally, the meaning of `equilibrium' appears intrinsically dependent on time. As Callen emphasizes: ``in all systems there is a tendency to evolve toward states in which the properties are determined by intrinsic factors and not by previously applied external influences." These are the equilibrium states, which are ``by definition, \textit{time independent}" (1985, 13) such that ``the properties of the system must be independent of the past history" (1985, 14).\footnote{Other textbooks make similar claims about the temporal nature of equilibrium. Buchdahl (1966) defines equilibrium via staticity -- lack of change over relevant timescales. Landau \& Lifshitz (1980) notes how equilibrium states are states which are necessarily arrived at \textit{after some relaxation time}. Caratheodory's (1909) discussion of equilibrium also focuses on relaxation time. Schroeder (2021, 2) introduces thermal equilibrium as such: ``After two objects have been in contact \textit{long enough}, we say that they are in thermal equilibrium." Matolcsi (2004) conceptualizes equilibrium via the \textit{standstill} property: a process is standstill when they are \textit{not varying in time} and have \textit{vanishing dynamical quantities}.} In other words, it seems almost \textit{a priori} that equilibrium is dependent on some background time, along which processes evolve, properties cease to change, and states terminate in quiescence. 

Hence, in the timeless context, we cannot simply claim that these equilibrium properties obtain. Furthermore, if equilibrium is defined in terms of time, and thermal time requires equilibrium, then thermal time doesn't solve the problem of time, since it takes time! \textit{Any} (faithful, normal) state can be deemed to be an `equilibrium' state with respect to its modular group's `thermal time', but this renders the meaning of equilibrium arbitrary. Instead, we need some story for why a state is `really' in thermal equilibrium. Such a story is typically provided with respect to some background time, and it's unclear how \textbf{TTH} can avoid referring to time in defining equilibrium. 

An immediate response would be to reject standard accounts of thermodynamic equilibrium. After all, those accounts were developed classically, without the problem of time in the background. To appeal to those accounts in the new context of the problem of time seems to be unfair. The question, then, is what notion of thermal equilibrium we should be appealing to, and what physical reasons we have for thinking that this new notion can recover our standard notions of thermodynamic equilibrium in domains where we already have a good grip on thermodynamic equilibrium.

\subsection{The time in timeless equilibrium}\label{3.2}

In response, Paetz (2010, $\S$7.6) suggests that we'd need an \textit{intrinsic} definition of equilibrium -- one that doesn't refer to time -- if \textbf{TTH} is to succeed. To my knowledge, the only noteworthy proposal is due to Rovelli (1993).

\begin{figure}
    \centering
    \includegraphics[scale = 0.3]{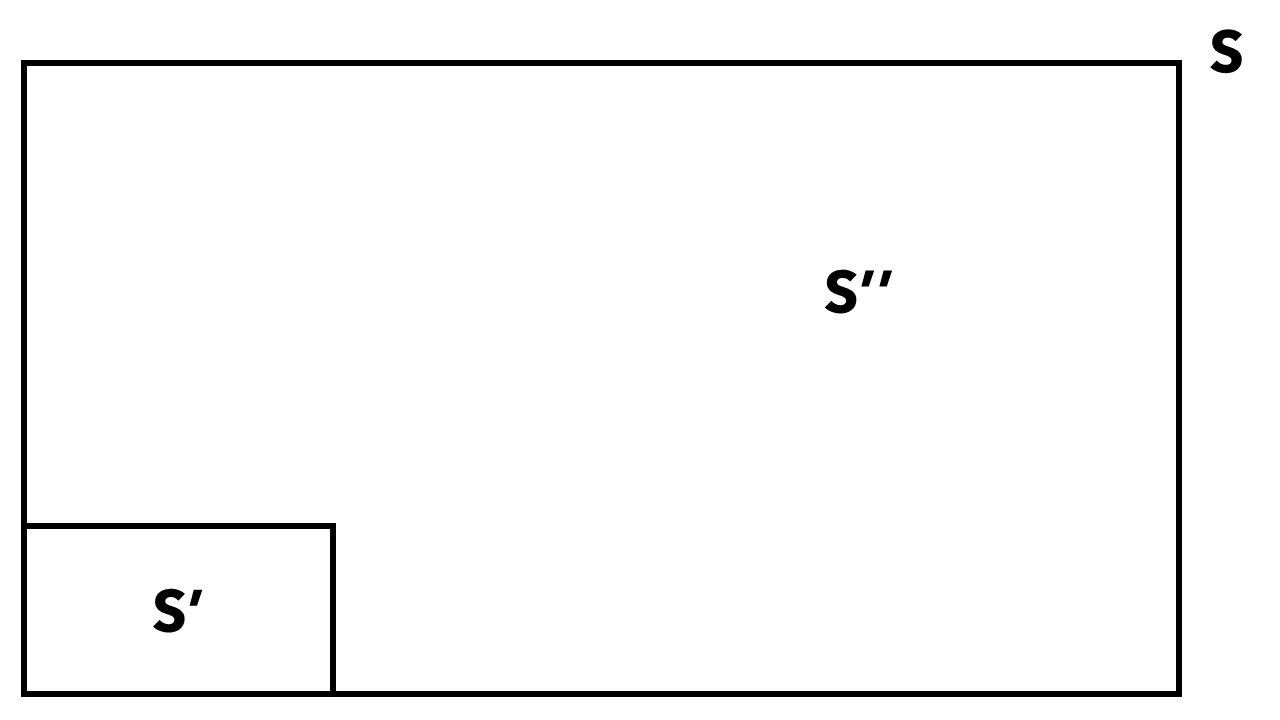}
    \caption{A partition of system $S$ into subsystems $S'$ and $S''$.}
    \label{fig:1}
\end{figure}

We can see how Rovelli's `timeless' definition of equilibrium is supposed to work, by seeing how it aligns with our standard concept of thermodynamic equilibrium in classical statistical mechanics. Rovelli (1993, 1559) claims that this condition was emphasized by Landau \& Lifshitz (1980) as a \textit{definition} of equilibrium.\footnote{To my knowledge, Landau \& Lifshitz does \textit{not} use this condition as a definition of equilibrium, but as a property which (more or less) holds for equilibrium systems.} For a system $S$ with phase space coordinates $p$, $q$, such that we can separate a small but macroscopic region $S'$ in spacetime, with associated coordinates $p'$, $q'$, from the (much larger) rest of the system $S''$ with coordinates $p''$, $q''$, and assuming weak interactions between $S'$ and $S''$, the interaction Hamiltonian approximately vanishes. (See Fig. \ref{fig:1}.) \textit{As a result}, for such a choice of $S'$ and $S''$, the probability distribution for the system -- its statistical state $\rho$ -- factorizes:
\begin{equation}\label{eq:21}
    \rho_S(p, q) = \rho_{S'}(p', q') \rho_{S''}(p'', q'')
\end{equation}
This condition essentially signals the statistical independence of one sub-system's statistical state from the other, and is prima facie free of time. One way to interpret this statistical independence is as representing a system being in equilibrium with itself by representing its parts (i.e. subsystems) as being in \textit{relative equilibrium with each other}. If these subsystems are in relative equilibrium, their thermodynamic properties will, of course, not change with respect to each other, and so it seems natural that the subsystem statistical states -- which determine macroscopic quantities -- are independent of each other and will factorize. Rovelli then proposes that this condition \textit{defines} equilibrium: ``we shall refer to equilibrium as a situation in which every small but still macroscopic component of the system is in equilibrium, in the usual sense, with the rest of the system." (1993, 1558--1559) 

Rovelli's proposal is supposed to extend beyond the instantaneous phase space of classical statistical mechanics into generalized phase spaces (compatible with the canonical approach to quantum gravity) where time is demoted, from a privileged, global, coordinate defining $p$ and $q$ at a time, to merely one of the many phase space coordinates. However, the idea remains exactly the same: if we can find systems whose parts are in relative equilibrium with each other (i.e. systems satisfying the factorizability condition for coordinates $p, q$), then those systems are in equilibrium simpliciter. For both classical and generally covariant contexts, this notion of equilibrium need not be global, i.e. need not hold true of all of phase space: to the extent that some large regions of phase space factorize per (\ref{eq:21}), that region can be said to be in equilibrium. If this works, then even in the globally frozen worlds described by the Wheeler-DeWitt equation, regions of these worlds can be ascribed thermodynamic equilibrium states, with which to define a (possibly non-global) thermal time via \textbf{TTH}.

To assess this proposal, let's precisify Rovelli's proposed definition. Firstly, a system is in equilibrium if and only if \textit{every} subsystem is in relative equilibrium with the rest of the system, viz. $S'$ and $S''$ are in relative equilibrium for all choices of $S'$ and $S''$ such that $S'$ and $S''$ are still macroscopic regions and $S'$ is significantly smaller than $S''$. Secondly, two subsystems $S'$ and $S''$ are in relative equilibrium if and only if \eqref{eq:21} holds.\footnote{Landau \& Lifshitz (1980, 7) notes that groups of subsystems also factorize with respect to the rest of the system in the same way, provided that these groups are still small enough relative to the rest of the system.}  

Unfortunately, I don't think that this definition of equilibrium is adequate. To begin with, the original physical justification for applying \eqref{eq:21} appears to rely implicitly on time, even if its form is explicitly timeless. What Rovelli does \textit{not} mention is Landau \& Lifshitz's caveat which immediately precedes \eqref{eq:21}: 
\begin{quote}
    It should be emphasised once more that this property holds only over \textit{not too long intervals of time}. Over a sufficiently long interval of time, the effect of interaction of subsystems, however weak, will ultimately appear. Moreover, it is just this relatively weak interaction which leads finally to the establishment of statistical equilibrium. (1980, 6, emphasis mine) 
\end{quote}
In other words, the application of this definition is manifestly justified in terms of background dynamics, just like other definitions of equilibrium. \eqref{eq:21} is clearly not intended to \textit{define} equilibrium states. Rather, the subsystems of systems in equilibrium can be justifiably characterized in terms of \eqref{eq:21} \textit{for suitable periods of time} but not \textit{always}. Relative to some timescales, subsystem interactions approximately vanish. Over long enough periods of time, interactions between subsystems, however small, render \eqref{eq:21} false. Macroscopic properties not changing for subsystems of a system in equilibrium does \textit{not} mean that their probability distributions, which depend on \textit{microphysical properties}, are likewise independent of each other. For all practical purposes, we may treat \eqref{eq:21} as approximately true, since we typically don't deal with systems on those time-scales. However, \eqref{eq:21} only holds true relative to certain timescales, and should not be taken to be a definition of equilibrium.

One possible response is to take Landau \& Lifshitz's definition but reject their physical justification. After all, they are clearly not working in the timeless generally covariant context, so, prima facie, we should not expect their justification to apply in this new context. Instead, we should treat \eqref{eq:21}, the factorization of statistical states, to define equilibrium for a generally covariant quantum system. Insofar as systems (approximately) factorize this way, we can take them to be in equilibrium and to define thermal time. We should therefore treat Landau \& Lifshitz's original physical justification -- that systems factorize \textit{because} they are weakly interacting subsystems in relative thermal equilibrium -- as a \textit{consequence} of this definition instead. \textit{Because} of this definition -- because systems approximately factorize in this way -- we can \textit{then} treat its subsystems as weakly interacting in relative equilibrium with each other. 

\begin{figure}
    \centering
    \includegraphics[scale = 0.3]{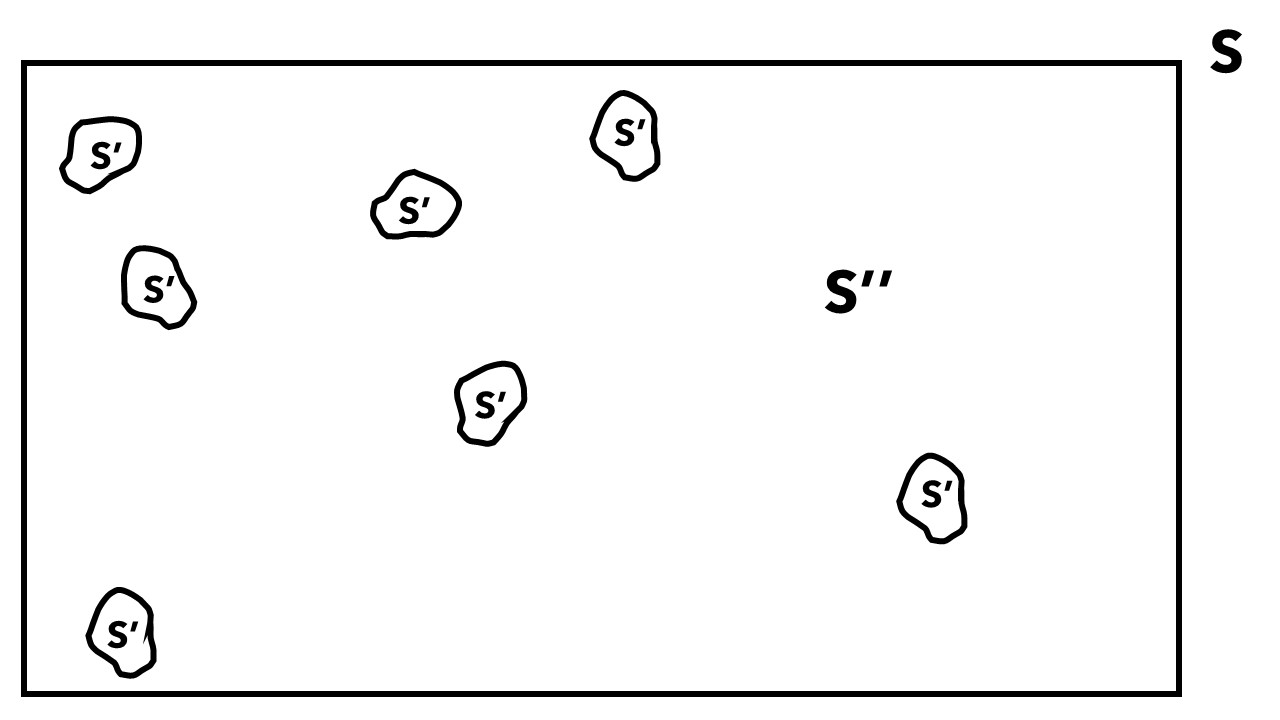}
    \caption{A schematic demonic partition of system $S$ into two subsystems $S'$, $S''$, where $S'$ is a composite subsystem of higher mean kinetic energy, and $S''$ is a subsystem of lower mean kinetic energy.}
    \label{fig:2}
\end{figure}

However, both parts of Rovelli's proposed definition encounter conceptual worries. Firstly, defining equilibrium in terms of relative equilibrium for \textit{all} choices of $S'$ and $S''$ is too strong. While it's true that a system would be in equilibrium if each subsystem is in such a relative equilibrium with the rest of the system, I don't think that the latter is \textit{necessary}, and hence cannot be definitional, for equilibrium. Even in the standard non-generally covariant context, any typical system, even those which we \textit{do} know to be in equilibrium, will not satisfy the criterion of relative equilibrium for \textit{all} choices of subsystems. The only requirement proposed by Rovelli is that $S'$ and $S''$ are macroscopic subsystems, and that $S'$ is much smaller than $S''$. Landau \& Lifshitz (1980, 7) notes that the same relation holds for groups of subsystems so long as the group remains small relative to the rest of the system. However, without further constraints, there are always going to be gerrymandered `Maxwell's demon' partitions of the system into two subsystems:\footnote{This need \textit{not} be an actual partition (using walls, membranes, etc.), and so sidesteps the question of whether Maxwell's demon is physically realizable.} a small disconnected collection of subsystems containing all and only the faster particles with higher momentum, $S_{fast}$, and a much larger region of the system containing all and only the slower particles with lower momentum $S_{slow}$.\footnote{See e.g. Hemmo \& Shenker (2010).} (See Fig. \ref{fig:2}.) It seems to me that nothing rules out the possibility of partitioning the system this way. It then follows that $S_{fast}$ is at a much higher temperature than $S_{slow}$ since the former has higher mean kinetic energy. So it seems that relative equilibrium doesn't obtain for such a partition of subsystems, even though we know that the system is in equilibrium overall. 

% Many thanks to Craig Callender and Eddy Keming Chen for suggesting this example in personal correspondence.

\begin{figure}
    \centering
    \includegraphics[scale = 0.3]{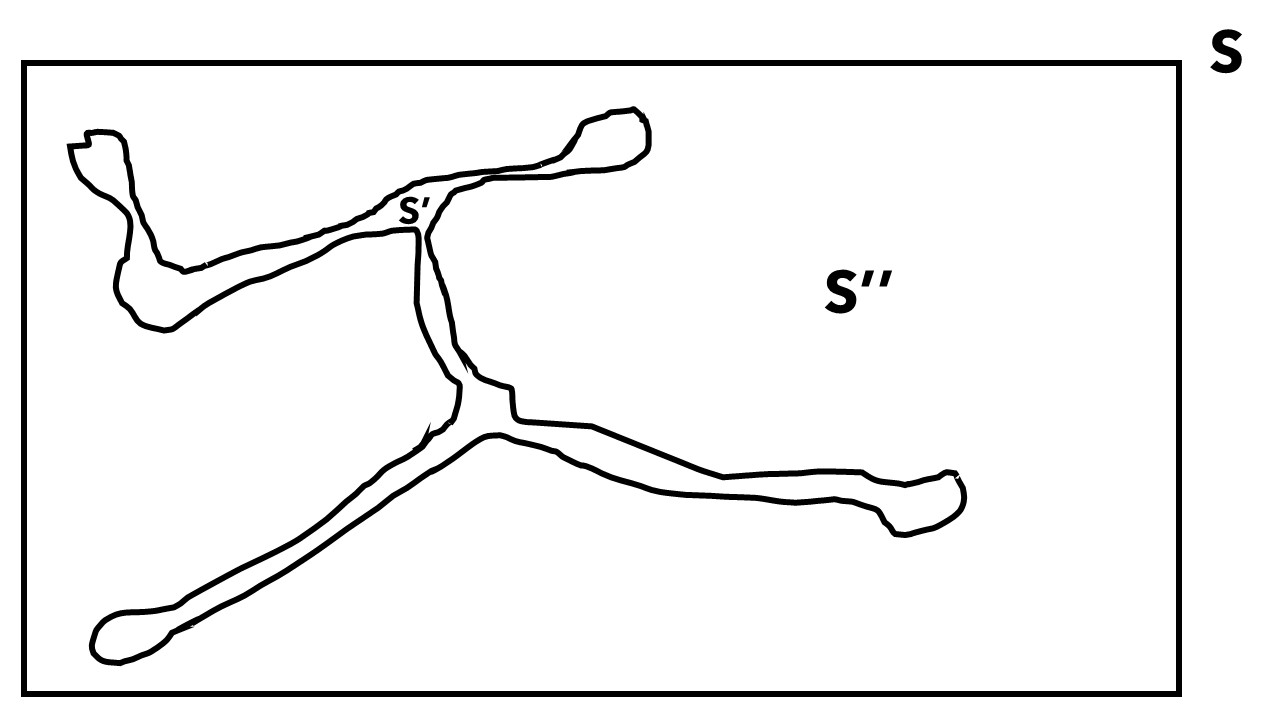}
    \caption{A schematic demonic partition of system $S$ into two subsystems $S'$, $S''$, where $S'$ is now connected but has higher mean kinetic energy. $S''$ has lower mean kinetic energy.}
    \label{fig:3}
\end{figure}

One reply to the demonic partition is to rule out disconnected subsystems. A revised definition of the first part of Rovelli's definition becomes: $S'$ and $S''$ are in relative equilibrium for all choices of $S'$ and $S''$ such that $S'$ and $S''$ are still macroscopic regions, $S'$ is significantly smaller than $S''$, and $S'$ and $S''$ are each \textit{connected} systems. This rules out (\ref{fig:2}) as an example, since $S'$ was a collection of disconnected subsystems. But this move does not exorcise the demon entirely, only the more visceral cases I raised above. Consider a new demonic scenario where $S'$ is a connected but highly gerrymandered subsystem which contains all of the high momentum particles but none of the lower momentum particles. (One can imagine this partition `weaving' through the space between particles to avoid just those lower momentum particles.\footnote{I owe this example to discussions with Craig Callender and Eddy Keming Chen.} See Fig. \ref{fig:3}.) On the contrary, $S''$ contains only the lower momentum particles. The same problem arises as before: the two subsystems are not in relative equilibrium yet the total system is in equilibrium ex hypothesi. So a demand of connectability would not resolve the worry I've just raised above.

\begin{figure}
    \centering
    \includegraphics[scale = 0.3]{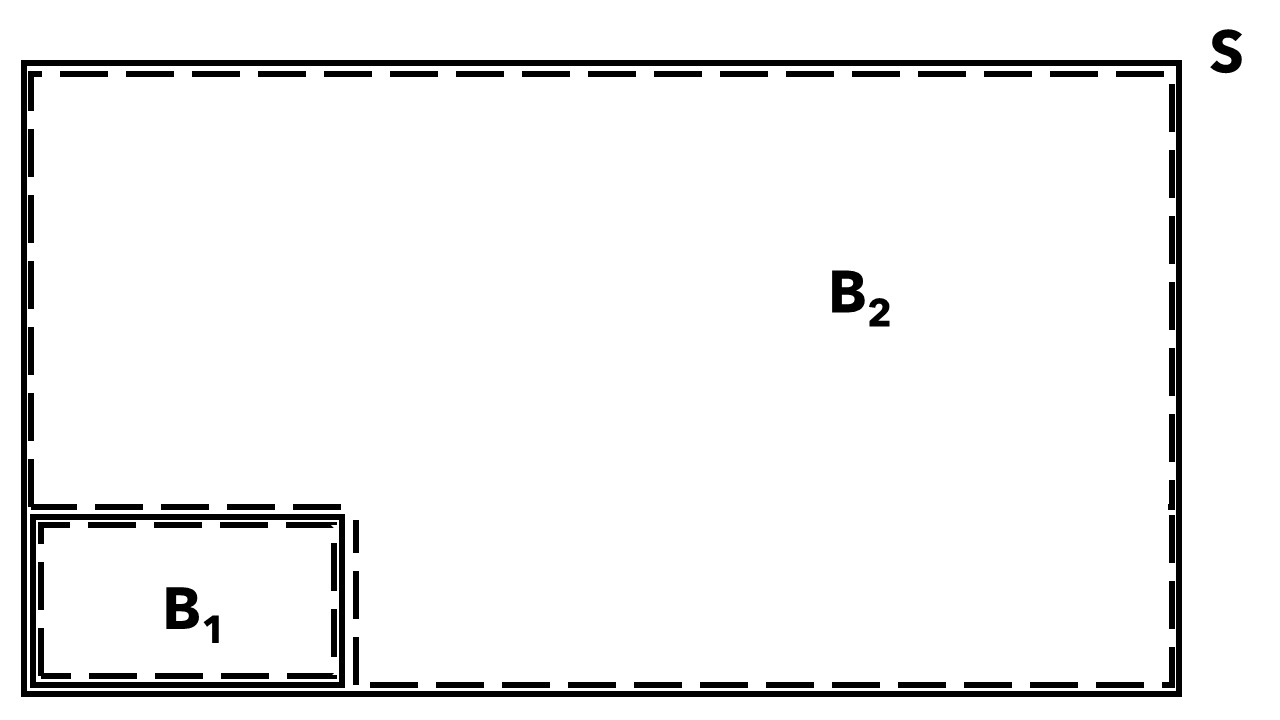}
    \caption{A system $S$ containing two (approximately) non-interacting boxes $B_1$, $B_2$, such that $B_1$'s temperature is not equal $B_2$'s. Their states factorize since they are non-interacting, but they are not in relative thermal equilibrium.}
    \label{fig:4}
\end{figure}

Secondly, there's a worry about whether mere statistical independence, i.e. \eqref{eq:21} -- the factorization condition -- suffices to track whether two subsystems are in relative equilibrium. Consider the simple case of a system of two boxes, $B_1$ and $B_2$. (See \ref{fig:4}.) $B_2$ can be much larger than $B_1$. The boxes are thermally insulated, electromagnetically shielded, and contains air at different temperatures. It seems to me that we can ascribe a statistical state $\rho_B$ to the joint system of $B_1$ and $B_2$, and that, at least for some regimes,\footnote{$\rho_B$ might be factorizable into $\rho_{B_1}$ and $\rho_{B_2}$ simpliciter if we have perfect thermal insulation and perfect mirrors preventing the transmission of radiation. Otherwise, there'll be some regimes for which we can ignore thermal radiation, and for which we might plausibly assume factorizability. For whatever regime in which factorizability holds, there's no clear physical sense in which the two systems are in relative equilibrium because they are not at the same temperature.} $\rho_B$ is factorizable into two subsystem statistical states $\rho_{B_1}$ and $\rho_{B_2}$, each describing the statistical state of the respective boxes. Taken as is, the proposed `timeless' definition of relative equilibrium appears to hold. However, it does \textit{not} suffice to characterize these two boxes as actually being in relative equilibrium: the two boxes are, ex hypothesi, at different temperatures. 

Again, one might be understandably tempted to reply that connectability -- this time \textit{between} subsystems, as opposed to within subsystems -- is the issue here. If we simply demand that the two subsystems must be connected, this would demand that we forgo the insulation barriers and such. Consider a revised version of the second part of Rovelli's proposal: two subsystems $S'$ and $S''$ are in relative equilibrium if and only if \eqref{eq:21} holds, \textit{and} $S'$ and $S''$ are connected. Then it seems prima facie plausible that two connected subsystems would be in relative equilibrium when their statistical states factorize. But one must be careful here and ask: what is it about \textit{connection} per se, that is supposed to be thermodynamically relevant here? Suppose for a second the scenario where $B_1$ will \textit{never} interact with $B_2$ no matter what, and the two boxes remain frozen at distinctly different temperatures, even when we removed the barriers. If the world were indeed truly timeless and frozen per the Wheeler-DeWitt equation, this could be a plausible scenario. In this scenario, though, it seems that the systems would indeed satisfy the factorizability condition: changes in one subsystem will leave the other subsystem unchanged. However, it seems wrong to say that they are in thermodynamic equilibrium since they remain at different temperatures. Of course, in everyday thermodynamics, $B_1$ and $B_2$ -- once connected -- would spontaneously arrive at the same temperature, and we would say then that they are indeed in thermal equilibrium. But note that connection here is merely a means to an end: it allows for -- but does not guarantee! -- \textit{interaction} between $B_1$ and $B_2$. Once we recognize this, though, the most plausible definition of thermal equilibrium appears to become: two subsystems $S'$ and $S''$ are in relative equilibrium if and only if \eqref{eq:21} holds, \textit{and} $S'$ and $S''$ are \textit{interacting}. But interaction is something that happens dynamically, and thus betrays Rovelli's proposal to define equilibrium in a timeless fashion.

\begin{figure}
    \centering
    \includegraphics[scale = 0.3]{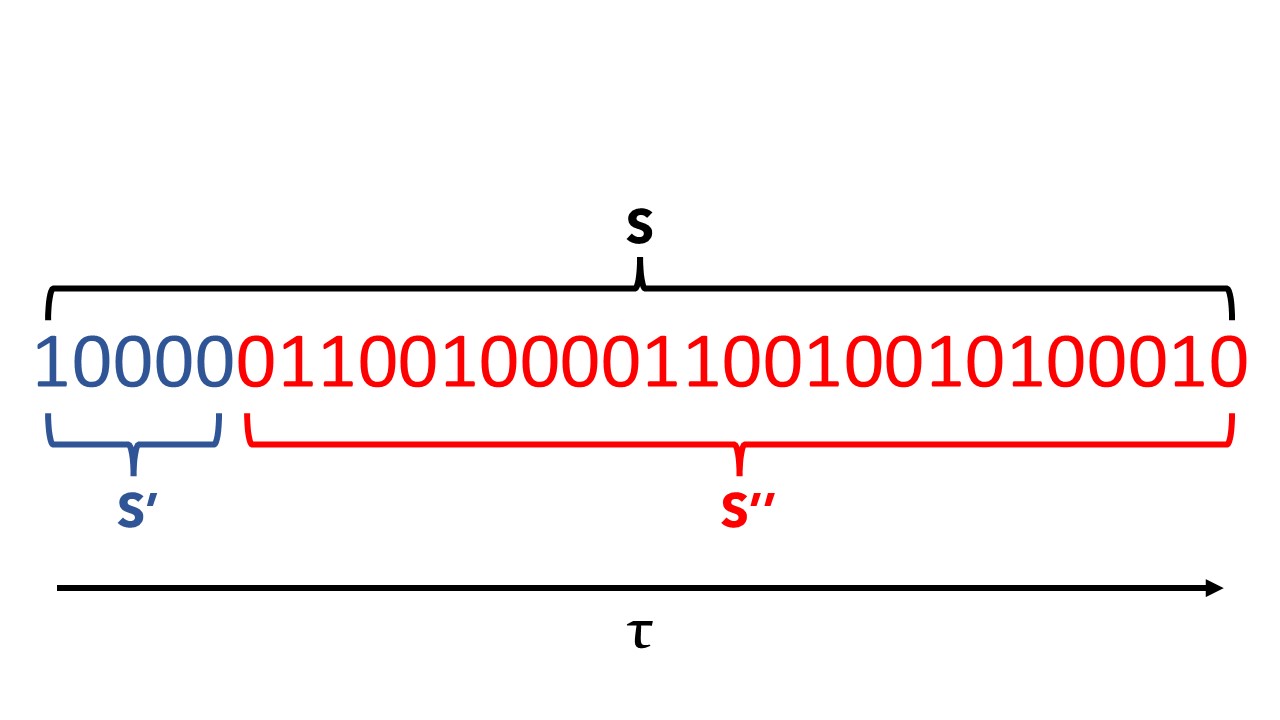}
    \caption{A probabilistic process with outcomes \{0, 1\} each with some probability of occurring every time-step $\tau$. Outcomes at each $\tau$ are probabilistically independent of future and past outcomes. The probability distribution for the sequence $S$ clearly factorizes for any choice of sub-sequences $S'$, $S''$.}
    \label{fig:5}
\end{figure}

This problem with using factorizability to entirely characterize relative equilibrium is furthermore amplified when we are not allowed to constrain our considerations to systems -- and states -- at a time, as when we consider generalized phase spaces and the context of the problem of time. Consider the application of \eqref{eq:21} to a system $S$ undergoing a probabilistic process such that at each time-step $\tau$, the state of the system at $\tau$ is either 1 or 0 with some probability. Furthermore, the state is probabilistically independent of future and past outcomes. Then, for any arbitrary partition of the entire sequence \textit{across} time into sub-sequences, the two `subsystems' factorize and so satisfy the timeless definition of relative equilibrium. (See Fig. \ref{fig:5}.) But it's clear that these subsystems are \textit{not} in relative thermal equilibrium -- they are simply probabilistically independent of each other. 

The above problems would not be troubling if we took the proposed condition to hold \textit{over time} and allowed the subsystems to \textit{interact}, viz. if the definition were justified by appeal to a background time. Then $S_{fast}$ would quickly lose energy to $S_{slow}$ and equilibrate over time. Likewise, the non-interacting subsystems would not be said to be in relative equilibrium, in virtue of not interacting with each other. Similarly, the different temporal parts of the probabilistic process will never interact with each other and hence are not in thermal equilibrium. 

But this move needs time and dynamics, returning us back to Landau \& Lifshitz's original physical justification. For Landau \& Lifshitz, \eqref{eq:21} is justified when we're allowed to take a system to be comprised of \textit{quasi-closed} systems -- when these subsystems \textit{interact weakly} with each other. (1980, Ch. 1, $\S2$) In the original context, the statistical independence encapsulated in \eqref{eq:21} seems to be justified in terms of (approximate) \textit{lack of interaction}. But interaction appears to be something steeped in \textit{dynamics} and thus time. It doesn't seem to be something we have without time. Since we're not allowed this justificatory resource, we're not allowed this natural solution. 

To sum up, Rovelli's `intrinsic' definition of equilibrium was originally justified by Landau \& Lifshitz with respect to some background time and dynamics. In the timeless context, we can of course reject Landau \& Lifshitz's justification on grounds of irrelevance. However, the definition, on its own, appears to be inadequate for defining equilibrium: relative equilibrium for all choices of partitions is unnecessary for characterizing equilibrium, and factorization is insufficient for characterizing relative equilibrium. Without an unproblematic timeless definition of equilibrium, the worries from before return: \textbf{TTH} cannot take off in a timeless context because we seem to require time to define thermal equilibrium as I've argued in \S3.1, and without the concept of thermal equilibrium, we cannot pick out the right kind of states $\mathcal{\omega}$ over $\mathcal{W}$ with which to define thermal time without circularity.

\subsection{The time in the modular group}\label{3.3}

From both the standard and the `timeless' thermodynamic perspectives, \textbf{TTH} seems to run into the same conceptual worry: to derive time from a frozen world, we must appeal to time to begin with via dynamical considerations. This problem can be seen more generally from the algebraic perspective as well. Broadly, the problem is this. The modular group doesn't necessarily have dynamical meaning. Yet, the dependence of \textbf{TTH} on the modular group requires interpreting the modular group as a bona fide dynamical object. However, any argument for justifying this interpretation seems to require appeals to prior dynamical considerations. Hence, any approach that relies on the modular group to derive time from no-time, including the \textbf{TTH} as a special case, seems to run into circularity.

To be clear, I am \textit{not} saying that the modular group cannot be formally defined as an abstract object. Mathematically, the modular group comes `for free' once we start with faithful, normal, states, thanks to modular theory. Recall $\S\ref{2.3}$: given faithful, normal states $\omega$, the Tomita-Takesaki theory guarantees the existence of a unique strongly continuous unitary group of automorphisms on $\mathcal{W}$, the modular group, parametrized by a single parameter $t \in \mathbb{R}$ such that:     
\begin{equation}\label{eq:27}
    \alpha_t^\omega A = \Delta^{-it} A \Delta^{it}
\end{equation}
Furthermore, $\omega$ is a KMS state with respect to the modular group. If we interpret $t$ to be time, then $\omega$ looks like a state in thermal equilibrium, remaining stationary along $t$. Conversely, if we interpret $\omega$ to be a thermal state, then $t$ is the time along which $\omega$ is in thermal equilibrium. 

I've already argued in \S3.1 and \S3.2 that the latter move cannot be made in a timeless context without running into circularity, since we have no time-independent way of defining thermal equilibrium and thermal states. For the same reason, the former move cannot be made in a timeless context. But it seems to me that this worry will generalize to \textit{any} attempt to interpret the modular group dynamically. 

To see this worry more explicitly, let us zoom out and look at the nature of the modular group. The modular group associated with a defining faithful, normal state is simply a strongly continuous one-parameter unitary group such that this faithful, normal state remains invariant under the action of the group, $\alpha_t^\omega$. Recall from (13) that
\begin{equation}
\omega(\alpha_t^\omega A) = \omega(A)
\end{equation}
Formally, a one-parameter unitary group is simply a group of transformations which conserves the inner product of Hilbert space $\langle\psi_1 | A | \psi_2 \rangle$ for states $\psi_1$, $\psi_2$, and operators $A$, under the action of one-parameter unitary operators. Wigner's theorem tells us that any symmetry of a quantum system corresponds to either a unitary or antiunitary group of actions on Hilbert space; a one-parameter unitary group can then be understood as implementing a one-parameter symmetry.

The most well-known one-parameter unitary group is of course the time-translation group associated with $e^{-iHt}$, where the parameter is the time $t$, which drives time-evolution under either the Schrodinger or Heisenberg equations given a physical Hamiltonian $H$. But, as I've already alluded to before, \textit{the time-translation group is not the only one-parameter unitary group}. Notably, thoroughly \textit{atemporal}, \textit{spatial}, symmetries such as rotational symmetry under some angle $\theta$ about some axis, spatial translational symmetry by some distance $x$ along some axis, or spatial scale symmetry parametrized by a scale-factor $\lambda$, can also be implemented as one-parameter unitary groups in principle. Furthermore, while these symmetries can be understood \textit{dynamically} as systems evolving from one symmetric state to another, they can also be understood \textit{passively} as redescriptions of the same system in terms of different symmetrically related coordinates.\footnote{This corresponds to the distinction between active and passive symmetries.} This is what I meant when I said earlier that the modular group can but need not be interpreted dynamically; in general, there will be many one-parameter unitary groups which are clearly \textit{not} dynamical depending on the symmetries of the system in question. Just because the modular group is a one-parameter unitary group does not mean it automatically matches the time-evolution given by the Heisenberg or Schrodinger equations. The modular group's physical meaning is in general underdetermined. 

Those who wish to interpret the modular group dynamically given this underdetermination of physical meaning -- e.g. to treat its associated modular `Hamiltonian' as a genuine physical Hamiltonian associated with dynamics -- must provide some physical argument for why we should do so. We've already seen one proposed solution to this problem, which is the conceptual core of \textbf{TTH}: if we start with a thermal state, then the modular group acquires natural meaning as the time along which a thermal state remains in equilibrium. But to define something as thermal requires time to begin with, as I've argued. It also seems to me that this worry will generalize: any attempt to resolve the underdetermination of the modular group's physical meaning must appeal to dynamical considerations in order to motivate why a natural interpretation of the modular group will be dynamical, but this runs us into circularity: we wanted to use the modular group to define time to begin with.  

\begin{figure}[ht]
    \centering
    \includegraphics[scale = 0.375]{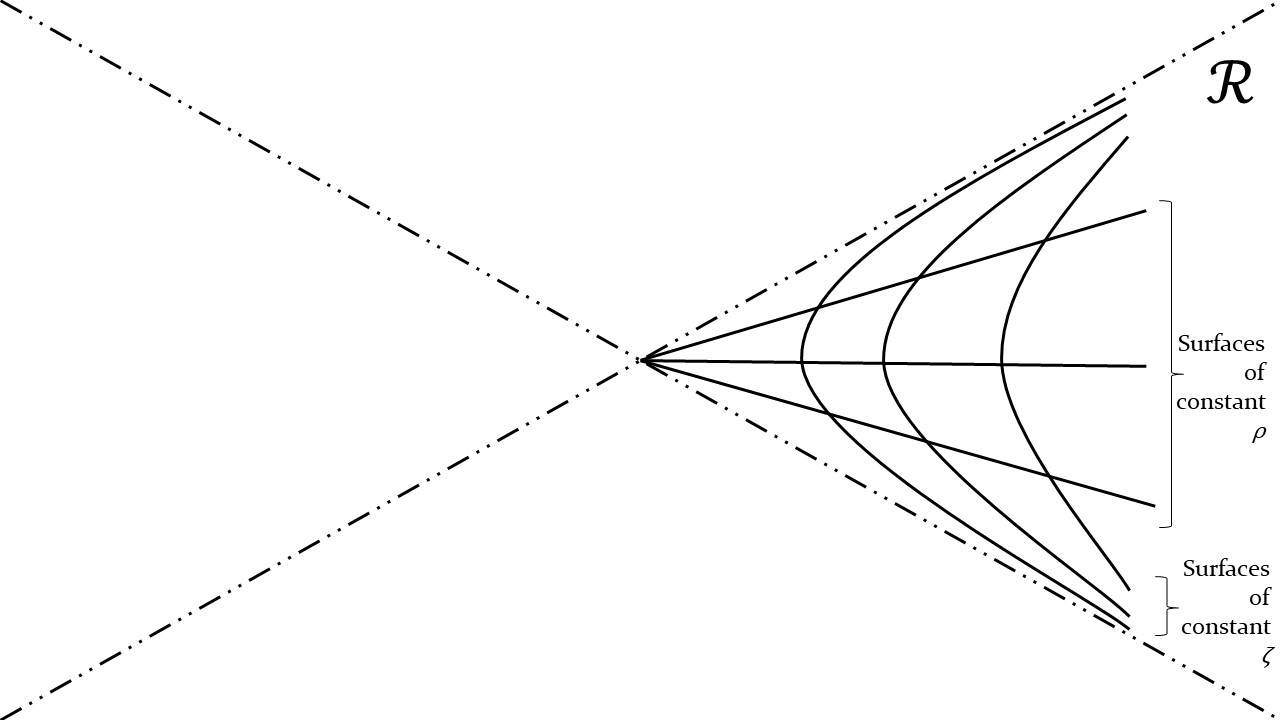}
    \caption{Schematic representation of the right Rindler wedge $\mathcal{R}$ in Minkowski spacetime. Rindler coordinates $\{\zeta, y, z, \rho\}$ are related to Minkowski coordinates $\{x, y, z, t\}$ by $x = \zeta \text{cosh} \rho$ and $t = \zeta \text{sinh} \rho$. The Minkowski metric in Rindler coordinates becomes $ds^2 = d\zeta^2 + dy^2 + dz^2 - \zeta^2 d\rho^2$. Surfaces of constant $\rho$ and $\zeta$ are labeled. $\rho$ is `Rindler time'. Timelike surfaces of constant $\zeta$ are the worldlines of constantly accelerating observers. Dashed lines indicate null directions.}
    \label{fig:rindler}
\end{figure}

Consider another related argument for when we can interpret the modular group dynamically outside of the timeless context: the case of immortal, constantly accelerating observers in the right wedge of Rindler spacetime. There are well-known results from Bisognano \& Wichmann (1975, 1976) suggesting a connection between the modular group and spacetime geometry in the relativistic context. Specifically, given a Minkowski vacuum state over the Weyl algebra $\mathcal{A}(\mathbb{R}^4)$ of the Klein-Gordon field and the associated von Neumann algebra $\mathcal{W}(\mathcal{O})$ associated with an open region of spacetime $\mathcal{O}$, the restriction of the algebra to the right Rindler wedge $\mathcal{R}$ (see \ref{fig:rindler}.) leads to a geometrical interpretation for the associated modular group for the Minkowski vacuum state: its generators are Lorentz boosts on $\mathcal{R}$.\footnote{See Earman (2011), Swanson (2021).} As Fredenhagen (1985, 79) explains, given the Bisognano-Wichmann theorem, the modular group's unitary operators $\Delta^{it}$ coincides as unitary operators $U(\Lambda_{2\pi t})$ implementing Lorentz boosts of the form: 
\begin{equation}
    \Lambda_{2\pi t} = \begin{pmatrix}
\text{cosh}2\pi t & \text{sinh}2\pi t & 0 & 0\\
\text{sinh}2\pi t  & \text{cosh}2\pi t & 0 & 0\\
0 & 0 & 1 & 0 \\
0 & 0 & 0 & 1

\end{pmatrix}
\end{equation}
In the right Rindler wedge, it's known that Lorentz boosts $\Lambda(a\tau)$ implement wedge-preserving time-translations along proper time $\tau$ for the orbit of observers with constant acceleration $a$. Since the modular group aligns with Lorentz boosts in this specific context, the modular group can be interpreted as a dynamical object; it aligns naturally with the proper time of constantly accelerating observers.

Crucially, this connection between the modular group and dynamics is again justified in virtue of the connection between the modular group and thermal states: the restriction of the vacuum state to $\mathcal{R}$ is a KMS-state relative to the modular group with Unruh temperature $\frac{a}{2\pi}$.\footnote{I set $c$, $k$ and $\hbar$ to 1. See Earman (2011).} This is the well-known Unruh effect: immortal, constantly accelerating observers in the right wedge of Rindler spacetime observes the vacuum state to be instead a thermal state. This provides a physical argument, again, for why the modular group aligns with the true dynamics of the system. Equivalently, the `modular' Hamiltonian aligns with the physical Hamiltonian in this special case.\footnote{Specifically, the former is proportional to the latter. See Arias et al (2017).} 

However, this connection between modular time and proper time exists only for a very specific class of models, and only for a very specific class of observers in said model: immortally and constantly accelerating observers in Minkowski vacuum. Generally, thermal time and proper time will not align. Swanson (2021) discusses the technical challenges that arise once we relax these assumptions. For instance, if we consider finite observers who has causal access not to the entire right Rindler wedge but a finite causal diamond (the intersection of their future light-cone at `birth' and past light-cone at `death'), the two quantities don't generally converge and modular time doesn't have the desired geometrical interpretation. Likewise when we consider nonuniform acceleration and nonvacuum states.\footnote{See Swanson (2021, \S3).}  In these cases, there is no dynamical justification, and hence no argument for interpreting the modular group dynamically. 

In short, for this particular case of the right wedge, we can resolve the underdetermination of the physical meaning of the modular group because, given certain conditions, the modular group action looks like something we already know to be dynamically relevant: Lorentz boosts. As Borchers \& Yngvason (1999) discusses, other cases for interpreting the modular group proceed similarly: for instance, instead of restricting attention to the right wedge $\mathcal{R}$ of Rindler spacetime, we can restrict attention instead to the forward light-cone region, in which case the modular group looks like the dilation group. If we restrict attention to the double light-cone regions instead, the modular group looks like the double-cone preserving conformal transformations. In these cases, we can interpret the modular group dynamically, but only because we already had a prior grasp of what the dynamics `look like'. 

So the situation looks like this: in general the modular group's physical meaning is underdetermined and does not necessarily have a dynamical interpretation. In each known case where the modular group \textit{has} a dynamical interpretation, it is by appealing to the existence of \textit{prior} dynamically relevant spacetime symmetries. This is of course no problem in the non-timeless context, but in the timeless context we should worry: if a dynamical interpretation of the modular group must appeal to prior dynamical notions, then the modular group cannot be used to define time and dynamics without circularity. This need not necessarily be the case, but to my knowledge, no other strategy exists. Proponents of the \textbf{TTH} will have to deal with this interpretative problem head-on if \textbf{TTH} is to take off. 

\section{Conclusion}\label{5}

To sum up, I've argued that \textbf{TTH} runs into a significant conceptual difficulty, in trying to explain how the modular group at the timeless lower level can fill the dynamical role played by time at the higher level. Without this explanation, time does not emerge: we've not explained why non-temporal structures like the modular group can be interpreted dynamically. However, in the known cases where the modular group can be interpreted dynamically, such an interpretation is typically motivated by appealing to a background (space-)time, implicitly or explicitly, as seen by the three perspectives discussed here. To secure time's emergence from the modular group, it seems, for now, that one must first start with time. Despite the elegance of the idea that time could be explained entirely by thermodynamic or algebraic considerations, \textbf{TTH} must overcome these worries in order to be a bona fide solution to the problem of time.

Importantly, I'm \textit{not} committed to the \textit{impossibility} of satisfactory justification. I simply want to emphasize that justification is \textit{presently} lacking for applying thermodynamic concepts, and dynamical concepts in the algebraic structure, in the timeless setting. I leave open the possibility that the challenges raised here may be met. \reviewerone{In particular, what is needed here is an argument for why the modular group can be interpreted dynamically, without appealing to background temporal or spatiotemporal concepts. To my knowledge, there isn't enough work on the \textbf{TTH} to support such a story for now.

One possibility, perhaps, is to develop in much more detail the idea that the modular group really is `time' in some weaker or diminished sense, even if doesn't generally play the dynamical role we associate with time at the higher level. On such a view, `time', in the form of the modular group, really fundamentally exists even in the supposedly timeless world, and can be understood to be `dynamical' in some sense. That is to say, quantum gravity might not be as timeless as we might think.\footnote{I thank an anonymous reviewer for suggesting this possibility.} However, only under certain conditions, as when the modular group aligns with the aforementioned conditions for interpreting it dynamically (in the usual sense, from the perspective of the higher level), does time-as-we-know-it emerge. But such a view would require motivating \textit{why} the modular group should be understood to be `time'-like, even as it doesn't possess all the typical characteristics of time-as-we-know-it. Furthermore, such a view would need to motivate how time might emerge beyond the very narrow set of conditions discussed in the previous sections, for instance, by generalizing thermal equilibrium and thermodynamics to a timeless setting, or generalizing beyond special sectors of Rindler spacetime for a very special class of observers. I don't see any \textit{a priori} reason why these are impossible, but there is still much more to be done in order to vindicate \textbf{TTH}.} 

In any case, this should make clear just how conceptually challenging the problem of time is for quantum gravity researchers. Many of the problems I have raised here are subspecies of more general problems of time plaguing quantum gravity.\footnote{See Anderson (2017) for a taxonomy of these problems.} A similar problem also arises in the semiclassical approach to the problem of time, where time is assumed to emerge approximately as a result of semiclassical approximations. Chua $\&$ Callender (2021) argues that these approximations, too, implicitly assume a background time for justifying their application, and are yet unjustified otherwise. There is ``no time for time from no-time", in their words. Likewise, here, it seems that there is, for the time being, `no time for thermal time from no-time'. 

\section*{Acknowledgments}

This paper was made possible through the support of a grant from the John Templeton Foundation. The opinions expressed in this publication are those of the author(s) and do not necessarily reflect the views of the John Templeton Foundation. I would like to thank Craig Callender, Eddy Keming Chen, Nick Huggett, Yichen Luo, Kerry McKenzie, Sai Ying Ng, Carlo Rovelli, Shelly Yiran Shi, as well as the anonymous reviewers of this manuscript (and earlier versions of it), for their very helpful remarks and comments that have improved this paper overall. I would also like to thank participants of the Lisbon-Lausanne workshop ``On Time in the Foundations of Physics", the International Summer Institute in Philosophy of Physics 2022, the Foundations of Physics @Harvard seminar series, the Philosophy of Time Society session at the 2023 Eastern APA, the Society for the Metaphysics of Science 2023 meeting, the European Philosophy of Science Association 2023 meeting, and the Caltech philosophy of physics reading group, for their questions and comments. 

\bibliography{bib.bib}
\nocite{*}
\end{document}